# Hybrid model for LTE Network-Assisted D2D communications

Thouraya TOUKABRI GUNES, Steve TSANG KWONG U and Hossam AFIFI

Orange Labs, Issy-les-Moulineaux, France
`(thouraya.toukabrigunes, steve.tsangkwongu)@orange.com`
Telecom SudParis, Evry, France
`hossam.afifi@telecom-sudparis.eu`

**Abstract.** In the evolution path towards the "Always-Connected" era and the trend for even more context-aware services, Device-to-Device communications (D2D) promise to be a key feature of the next-generation mobile networks. Despite remaining technical issues and uncertain business strategies, D2D-based services represent a new market opportunity for mobile operators that would manage to smoothly integrate these new technologies as a complement or even an efficient alternative to cellular communications. Existing research efforts on the integration of D2D technologies in cellular networks have mostly failed to meet user expectations for service simplicity and reliability along with operator requirements regarding lightweight deployment, control and manageability. This paper proposes a hybrid model for D2D communications assisted by mobile operators through the LTE network: it includes a lightweight D2D direct discovery phase and an optimized data communication establishment for proximity services. The proposed hybrid model is appraised against the existing solutions in literature and the current standardization effort on Proximity Services (ProSe) within the 3GPP.

**Keywords:** D2D communications, LTE networks, Proximity Services.

## 1 Introduction

Today's mobile networking world facts are: the mobile industry is shipping more smartphones and tablets than PCs; success stories of social networking services like Facebook has become a social trend from which mobile users developed the need to be connected anywhere to their surroundings; statistics in [4] envision an exploding number of more than thousand billion wireless connections around the world in 2020. Meanwhile, revenues of mobile services have been growing at a much slower rate than the growth of mobile connections since 2011[1]. The challenge for mobile operators, who struggle each quarter to turn a profit on voice and SMS services, is yet to face the threat of Over-The-Top (OTT) providers who have put their foot down at the mobile market with apps that supply instant messaging, multimedia services like photo sharing and video conferencing and other popular services for free. In this context,

---
[1] http://www.wwpi.com/

Device-to-Device communications have become the new driver in wireless networking and mobile market.

Defined as a short range direct communication between devices without the involvement of the network infrastructure, D2D communications have been proposed as an underlay to cellular networks. Such a solution will evolve cellular networks toward a layered topology in which multiple network layers (femto-network, D2D-network, Wifi-network…) would coexist with a main macro-cell layer.

With these new types of communications mainly based on context and proximity information, a new generation of user-centric mobile services will rise, offering at the same time the opportunity for operators to extend their mobile networks' capacities and to alleviate the traffic in their core networks; for instance, smart cities services, real-time social discovery of nearby persons, targeted and personalized hyper-local services (advertising, couponing/ticketing, restaurant/hotels booking, content download, etc.). Besides, when including group communications and relay mechanisms, D2D communications could be a relevant fallback alternative for the public safety services (police, firefighters, emergency services, etc…) in disaster situations (earthquake, Fire, etc.): using a specific D2D-enabled Public safety device, an officer/agent can exchange data and transmit information to other devices through a D2D group communication. Moreover, in poor radio coverage areas, relay-based D2D mechanisms could be an efficient way to extend network connectivity.

Surfing on the wave of the successful worldwide launch of 4G LTE (Long Term Evolution) mobile networks, a new short range technology based on LTE (LTE Direct) has been developed by Qualcomm[2]. Envisioned to be the next trendy D2D technology that best meets the requirements of the above mentioned types of services and successfully tested in a recent research work[3], 3GPP (3rd Generation Partnership Project) has then initiated a standardization effort on the integration of LTE Direct in mobile networks [12]. If this effort first addresses Radio Access Network (RAN) requirements and technical issues for the support of D2D-based services, it comes also along with a feasibility study [5] and a technical specification on the architecture enhancements for the support of Proximity Services (ProSe) [6].

In literature, many research works have been done on D2D communications and their integration within LTE networks. The earliest ones addressed mainly the radio aspects such as D2D radio interference management with cellular communications, power control, radio resources allocation/sharing methods and spectrum regulatory aspects (use of a licensed or unlicensed band for D2D). Studies have also been made on D2D discovery and communication mechanisms. However, the few proposed solutions in these fields are still immature and don't answer basic user concerns for simplicity, reliability and QoS when using D2D-based services. Otherwise, if these requirements have been answered by the recently standardized solutions proposed in 3GPP [6], the current specification is globally lacking from a more extensible and evolutionary vision of the D2D integration in current and next generation networks.

---

[2] http://www.qualcomm.com/solutions/wireless-networks/technologies/lte/lte-direct

[3] http://english.etnews.com/internet/2909211_1299.html

This paper proposes a novel LTE-based D2D discovery and communication mechanism. The solution aims to integrate D2D as native features in the current LTE architecture in order to promote the new paradigm of decentralized and locally-scoped communications. It mainly includes a reliable Direct Discovery phase and an efficient and optimized communication establishment phase. In the following sections, we give an overview on D2D use cases classification and Proximity Services (ProSe) as well as a discussion on existing D2D mechanisms. Then, we describe our D2D hybrid approach for discovery and communication and compare it to the current 3GPP standardized solution.

## 2    Overview on D2D use cases

Similarly to location-based communications, a D2D communication benefits from the proximity of devices in order to establish a direct link between them for a local data exchange. Basically, devices could be any device equipped with a D2D technology suitable for short range communication such as smart phones, tablets, laptops, network printers, cameras, or even connected vehicles.

When dealing with "devices proximity", the definition of "proximity" becomes questionable and different proximity levels could be defined (i.e. geographical proximity, network topology proximity (e.g. devices within the same subnet), radio range, etc.). For mobile operators with already a rich amount of valuable user data, context and proximity information are assets that need to be empowered to offer new value-added services to users and apace mainstream the adoption of D2D-based services.

**Table 1.** D2D use cases classification

| Use case category | Applications |
|---|---|
| **Commercial and Social Proximity Services:** An evolution of LBS services through hyper-local and dynamic proximity data. | • Discovery-centric services: Context-aware applications, Social networking applications, location enhancement applications, Social gaming, and smart cities services… <br> • Communication-centric services: content and video sharing services |
| **Public safety services based on group and relay communications:** Secure services used on specific D2D enabled devices and deployed on a dedicated non-public network | • Direct communication between public safety agents in or outside network coverage: push-to-talk, group communication, priority handling… <br> • Dedicated network access sharing for out of coverage devices through peer-to-peer connections to nearby in-coverage devices. |
| **Services for network capabilities enhancement** | • Offloading services: offload of local data traffic or video/voice call traffic. <br> • Multi-hop access services: Internet connection sharing through devices acting as relays, Connectivity extension to heterogeneous networks (UE acting as a gateway to Sensor network (MTC), UE in a vehicle acting as a cooperative relay to an ITS network infrastructure, etc.) |

### 2.1 D2D use cases classification

D2D-based services is a business opportunity for operators and present multiple attractive use cases going from public/commercial services to more specific fields like public safety and military services. Generally, D2D applications aim at proposing and improving local services for users while optimizing the throughput over the radio network, reducing the overall core network load and enhancing connection delays. Multiple use case classifications were made in [2], [5] and [6]. Generally, D2D-based services can be classified into three main categories like described in table 1.

### 2.2 Operator role in D2D

D2D communications are definitely a business opportunity for mobile operators. Operators' strengths consist in having a powerful network infrastructure that will allow the deployment of new D2D-based services while assuming their original contractual duty on keeping data security and users' privacy. Despite the set of technical issues and challenges that operators may face with the integration of D2D in their LTE networks, they still have rich valuable assets that meet user expectations toward new D2D services. Operators' important assets include:

― Service security and QoS through secure and uninterrupted connections.
― Identity management, authentication and Privacy when using a D2D service.
― Context information exposure for more attractive services and better QoE.
― Devices management: user's cellular and non-cellular devices are associated to his user profile and included into the operator's subscriber database and automatically associated with the owner's cellular devices.
― Ensure the consistency of the user experience including reachability and mobility aspects (seamless offload, seamless handover, etc.).

## 3    D2D peer discovery and communication establishment

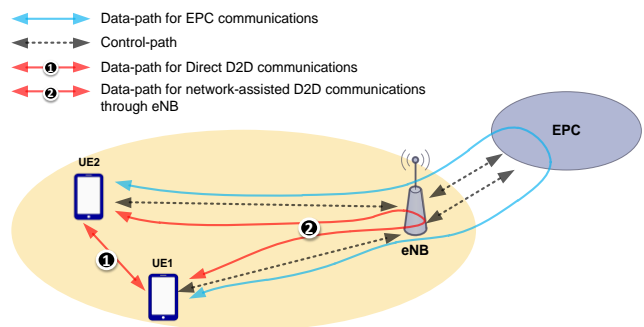

**Fig. 1.** Data path in direct and network-assisted D2D communication

A D2D communication has mainly two phases: A discovery phase in which devices aware of their location detect surrounding devices/services, and a Communication phase in which D2D peers exchange or share data on a D2D link.

Two communication modes were identified in [1], [2], [3] and [6]: the D2D Direct mode or the D2D network-assisted mode. As both D2D discovery and communication phases could be relatively independent, different models can be derived from each D2D mode according to the role of the operator in D2D phases. In a network-assisted D2D mode, two schemes could be derived: a fully-controlled scheme in which the operator have a full control on discovery and communication phases, and a loosely-controlled scheme in which discovery and communication phases are partially assisted by the operator, for instance with authentication mechanisms and radio resource allocation.

### 3.1 D2D peer discovery approaches

D2D peers need to discover each other before initiating a D2D communication. This discovery phase could be done directly between devices in an ad-hoc manner or with the support of an operator network that can either control the entire discovery phase by detecting D2D candidates at the core network level or only assist it as a trusted third party. Two main discovery approaches were identified [1] [2] [3]:

- Direct discovery approach:

According to 3GPP, D2D direct discovery is defined as the process of detecting and identifying devices in proximity using E-UTRA[4] direct radio signals [6]. Two discovery models were identified in ProSe: the "I'm Here" model (A) and the "who is there? / are you there?" model (B). In model A, a UE (User Equipment) could be an Announcing UE that broadcasts some information to its surrounding at pre-defined discovery intervals, or a monitoring UE that monitors certain information of interest from devices in proximity. In model B, the discovery model is more accurate about what is exactly needed to discover and thus, defines two roles for the UE: a "discoverer UE" which sends certain information about what it is interested to discover and a "discoveree UE" which replies with some information to the discoverer. The Direct discovery approach has the advantage of flexibility and scalability as it can adapt to an increasing number of D2D connections and allow in this way the offload of core network traffic to local D2D communications. However, such a method may have an impact on the UE complexity as it needs to support power management mechanisms to avoid battery consumption issues caused by an "always-on" discovery. Besides, specific radio resource scheduling and allocation mechanisms need to be defined in order to overcome interference issues with cellular communications. Moreover, this distributed approach may suffer from the lack of security and authentication mechanisms that need to be deployed in order to meet users' concerns about privacy.

- Centralized discovery approach:

This approach involves at least one or more network entities in the discovery procedure. As the operator network has a wider vision on the overall traffic and on UE mobility context, centralized discovery approaches aim at exploiting mobile operator core network assets about devices micro/macro mobility in order to provide a more accurate and efficient discovery information. The control of the discovery phase at the core network ensures the delivery of a better QoS and a more reliable service, thanks to authentication and privacy mechanisms offered natively by the operator. Centralized discovery is defined as EPC-level discovery in the 3GPP [5]. However, this approach could be less scalable than the direct approach as it could have performance and overload impacts on the core network with the additional D2D traffic. Efficient load-balancing and lightweight signaling mechanisms should be implemented at the core network in order to make D2D profitable for the operator and the user.

---

[4] E-UTRA: Evolved UMTS Terrestrial Radio Access

### 3.2　D2D communication establishment

After the discovery phase, D2D peers establish a communication link for data exchange. As shown in figure 1, the exchange of data could be done either directly between D2D devices, or using an optimized path through the eNB.

Basically, the setup of an EPC-based communication [8] [11] [13] consists on the establishment of an EPS (Evolved Packet System) bearer composed of a radio bearer (E-UTRAN bearer between UE-eNB), an EPC bearer and packet filters. There are two types of EPS bearers: the default bearer and dedicated bearers. At the initial network attach of the UE, an EPS default bearer is setup by the establishment of a PDN (Packet Data Network) connection with the PDN Gateway and the allocation of an IP address to the UE. In the case of a D2D communication, a dedicated D2D bearer with specific radio resources allocated by the eNB is setup. However, there is no need to setup an EPC bearer as the data path is optimized for D2D communications. Since the D2D UE has already an IP address associated with its default bearer, it is used for all bearers within the same PDN connection including the D2D dedicated bearer. Based on the concept of QoS-based flow aggregates, multiple D2D flows corresponding to simultaneous D2D connections could be carried within the same D2D data radio bearer.

According to the current 3GPP standard on ProSe [5] [6], direct D2D communications are allowed only when devices are out of coverage and only for Public safety services and bearer mechanisms are not yet defined for these types of communications. In [8], a D2D dedicated bearer mechanism is proposed for a D2D offloading service. Generally, existing D2D literature lacks from detailed mechanisms for D2D bearer establishment for D2D-based services other than public safety and offloading.

### 3.3　Existing solutions

Several solutions were proposed in D2D literature for D2D discovery and communication establishment mechanisms. They are summarized as follows:

- Solution 1 [3]: an EPC-based D2D discovery through MME using the Session Initiation Protocol (SIP). A SIP dedicated handler is proposed as an extension the MME in order to process D2D SIP packets. D2D SIP packets are encapsulated in NAS packets. The main limitation is the protocol layer violation between the User plane and the Control plane when processing SIP messages in the MME.

- Solution 2 [3]: an EPC-based D2D discovery through gateways for offloading. A D2D filtering is implemented in the gateways to discover D2D candidates. All IP flows associated to the same tunnel endpoints are filtered and marked as D2D. Radio measurements are made by eNB to decide to offload or not the traffic. The deployment of such a solution would have an impact on EPC entities performance (congestion, resource consuming).
- Solution 3 [8]: an EPC-based D2D discovery through a ProSe function implemented at gateways for offloading. A D2D dedicated bearer control mechanism is proposed. The limitation of this solution is that the dedicated bearer setup procedure is done in EPC while the data exchange is done directly between devices. Also, the additional IP address allocated for the dedicated EPC bearer would break the session at the handover from EPC to D2D link.
- Solution 4 [6]: a Direct D2D discovery through a dedicated ProSe server. A ProSe authorization mechanism is proposed before the direct discovery. The Direct discovery is made over IP through a ProSe protocol that is to be defined in [12]. D2D bearer control procedures are not yet defined and ProSe communication mechanisms are specified for one-to-many schemes only.

**Table 2.** Existing D2D solutions comparison

|  | Solution 1 | Solution 2 | Solution 3 | Solution 4 |
|---|---|---|---|---|
| **Direct or EPC-based discovery?** | EPC-based discovery | EPC-level discovery for offloading | EPC-level discovery for offloading | Direct discovery |
| **D2D bearer mechanism?** | No | No | Yes | No |
| **EPC impacted entities** | MME | PGW, SGW, eNB | SGW, PGW, MME, eNB | ProSe, HSS, MME |
| **Support for D2D Authorization?** | No | No | Yes | Yes |
| **Support for session continuity at handover?** | No | No | No | No |

As described in table 2, most of the existing solutions propose EPC-based discovery mechanisms with weak contributions on the communication establishment mechanisms. D2D authentication and authorization are not seriously considered except at the 3GPP ProSe solution. Besides, IP session continuity at handover from EPC-based communications to D2D is not yet discussed in any of the above-mentioned solutions. Generally, the proposed LTE improvements to support D2D communications are lacking from a long term vision to the viability of the EPS architecture: adding complexity to a system without considering its evolution in the future and regardless to economic and scientific advances may lead to its fast obsolescence. In next sections, we expose a long term vision of the deployment of D2D communications through a hybrid LTE network-assisted D2D solution for Discovery and communication.

## 4 A hybrid model for D2D discovery and communication

### 4.1 Motivations and proposal

The main motivation for D2D-based services was originally the offload of local communications through an optimized data path (D2D path) in order to alleviate the traffic overload in the core network. With the expansion of context aware applications market, the motivation has evolved toward a new type of mobile services based on proximity information. In the vision of future cellular networks [9], studies have demonstrated the tendency to move toward more flat architectures in which core network functions are virtualized in order to enhance the performance and reduce operational costs of core network extensions for operators. In this context, Cloud-Radio Access Network (C-RAN) has been proposed as an enhancement to the LTE RAN. It consists mainly on the optimization of eNB deployments through the virtualization of some of its functions. Composed of a BBU (Baseband Unit) and a RRH (Remote Radio Head), the eNB BBU function is virtualized and centralized [10] in order to reduce deployment costs in urban areas, efficiently use processing resources, limit inter-eNB interference issues and improve scalability on the RAN through collaborative radio mechanisms. In the following, we introduce a beyond-LTE vision of the D2D integration within the current EPS architecture.

### 4.2 Enhanced eNB role for local D2D services

From a network design point of view, the closest network entity to UEs is the eNB. Being the anchor point between the RAN and the EPC, the eNB disposes of location information of each UE in the cell (for devices in the active mode). As such, we propose to integrate main D2D functions in the eNB. The solution is based on a hybrid approach that combines a lightweight D2D Direct discovery mechanism assisted by the operator for the Authentication and Service Authorization aspects, and an optimized and QoS-enabled D2D communication through the eNB. Besides, we choose to minimize the involvement of EPC entities (MME and HSS) in order to avoid too much overhead in EPC. We also propose some D2D extensions to the LTE protocol stack and to implement the following functions as an evolution to the current eNB implementation:
  − Locating and checking D2D peers positions in cells (for devices being in the Active mode).
  − Identification of an active D2D communication.
  − Authorization and allocation of dedicated D2D discovery resources for the D2D pairs.
  − Mode selection by the means of specific radio measurements on the cell (global congestion check on the cell) and with the UEs.
  − Dedicated D2D Radio bearer establishment with D2D peers and resource allocation for the D2D communication.

### 4.3 A reliable Direct Discovery mechanism

Before initiating the D2D procedure, we assume that D2D enabled UEs have registered to a specific D2D service through a service platform. A D2D service registration phase consists basically on granting access credentials for a specific D2D service from an Application Server (AS) using a client application embedded on devices. From an LTE point of view, the service authentication could be done through the IMS (IP Multimedia Subsystem) if the service is IMS-based or directly between the UE and the AS through some authentication services offered by the operator (i.e. GBA[5]) if the service is IMS-independent. The proposed solution in this paper is agnostic to whether the service is IMS-based or not and the D2D service registration mechanism is out of this paper's scope. As depicted in figure 2, our solution is based on the following steps:

- Initial NAS Attach, D2D Authentication & Authorization:

In order to start a D2D Discovery, D2D peers need to exchange their identities and to be authenticated by the operator network as trusted D2D enabled devices. In order to relieve identity generation and management, we propose to use the Single Sign-On Authentication concept (SSO)[6]: each D2D enabled UE has a unique D2D identity (D2D_id) stored locally in the UE. Such unique identifier is also mapped to the user IMSI and is stored among the subscriber profile information at the HSS. Basically, at power on, a UE initiate a Network Attach procedure in order to beneficiate from the NAS-level services, after which a UE context is setup at the MME, a default bearer is established between the UE and the PDN-GW and a UE IP address is allocated. As a pre-discovery phase and in order to secure the discovery between D2D enabled UEs, we propose to include a D2D authorization procedure during the network attach: the IMSI[7] is sent into the **NAS_attach_req()** message to the MME with an indication to check the D2D_id during the authentication phase with the HSS. An **Update_location_req()** message is then sent by the MME to the HSS in order to get the subscriber profile information [7]. The D2D_id is then retrieved from the HSS user profile by mapping with the IMSI together with the list of D2D authorized services for that specific UE. This D2D authorization information is then stored at the eNB on a per UE D2D_id basis and is refreshed at the expiration of a validity timer.

- PDN connection & default bearer establishment:

Once the D2D UE is authenticated and authorized, the default bearer is created through the PDN connection and an IP address is allocated to the UE. This bearer is maintained during the D2D communication in order to ease the handover to the EPC path if the D2D connection is broken. Moreover, as the UE does not change its PDN connection while initiating a D2D communication, the same IP address allocated at the initial NAS attach is used by the D2D dedicated bearer described hereafter.

---

[5] http://en.wikipedia.org/wiki/Generic_Bootstrapping_Architecture
[6] http://en.wikipedia.org/wiki/Single_sign-on
[7] IMSI: International Mobile Subscriber Identity

- Resource allocation for direct discovery:

Assuming that a D2D application is used at the UE to access a specific D2D service, a specific D2D request message is sent from the UE to the eNB. The **D2D_direct_discovery_request (UE D2D_id, D2D_app_id)** is an extented RRC message that is sent to the eNB to request the authorization to use direct discovery resources. The packet is processed at the eNB which checks the authorization with the MME for the specified D2D_id in order to allocate the resources for the direct discovery over the E-UTRA. Once the authorization for direct discovery resource allocation is granted, the eNB generates a temporary D2D_id to the requesting UE for the direct discovery purposes and sends back an acceptance notification to the UE with the temp_D2D_id, as well as the needed configuration information for the authorized and allocated discovery resources (uplink/downlink, transmission power, etc.). The D2D generated temporary identifier is used to identify an active D2D connection at the eNB and expires with a validity timer.

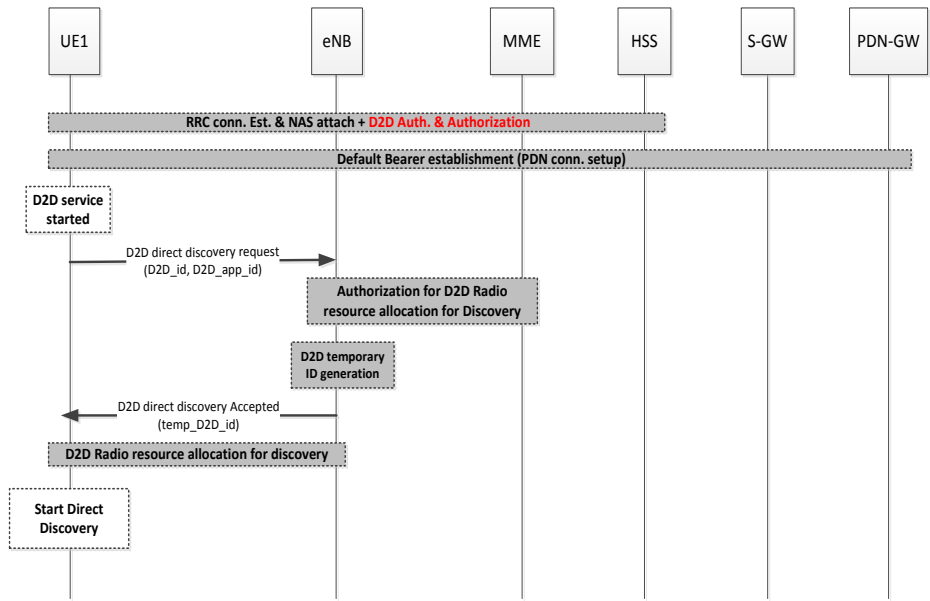

**Fig. 2.** D2D control signaling for a direct discovery over E-UTRA

- D2D direct discovery over E-UTRA:

After granting the authorization and the temp_D2D_id for discovery, the UE sets up through the D2D application its discovery mode according to the discovery models described in section (3.1). Proximity discovery messages are then sent over the air to discover devices of interest.

### 4.4 An optimized and QoS-enabled communication through eNB

After the direct discovery, a data path should be setup. In the specific case of a D2D communication, the communication is optimized and the data path could be established whether directly between devices or through the eNB which plays the role of a network anchor for the communication. In both schemes, the establishment of an additional EPC bearer is not needed as the data are exchanged locally between the UEs.

In our proposal, we choose to have a controlled communication between D2D peers by using an optimized data path through the eNB. For the operator, charging and legal interception rules would be facilitated when the data passes through the eNB; this avoids implementing additional complex mechanisms on the UE to support such functions. In addition, the control of the D2D bearer establishment through the eNB allows adapting QoS parameters according to the D2D service type. It also maintains the session continuity; even if one of the D2D UEs is not anymore in the D2D range, the session is not broken as long as the D2D communicating UEs are in the coverage range of the eNB. Instead, a seamless handover would be done in order to bring back the D2D data path to normal EPC communication path. The concept of D2D dedicated bearer is explained in figure 3.

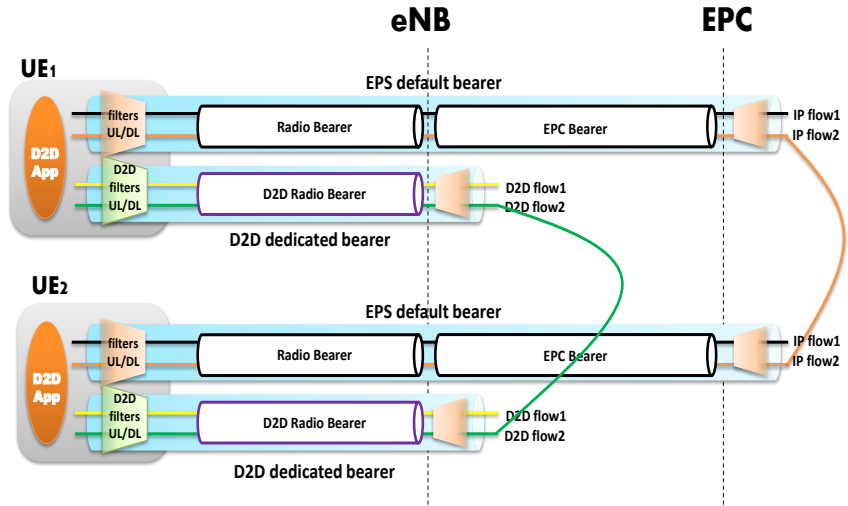

**Fig. 3.** D2D dedicated bearer concept for D2D communications meaning an optimized data path through the eNB

In figure 4, we propose a signaling call-flow for the establishment of a D2D communication after a discovery phase. Both discovered UEs send a D2D direct communication request to the eNB with their respective D2D_id. At the reception of these requests, the eNB will ask the MME for the establishment of a dedicated D2D radio bearer with each UE. The role of the MME here is to maintain a context for active D2D bearers in order to ease the handover to normal EPC connection when a D2D connection is broken. Each D2D dedicated bearer is identified at the MME by a D2D_bearer_id. A D2D bearer context is a D2D bearer in which an IP flow is routed. As explained before, there is no need to allocate a new IP address for the D2D dedicated bearer: using the same initially allocated IP address, the IP flow is routed on the D2D dedicated bearer through the eNB using a routing rule setup on the UE: the D2D packet filter is mapped with the correspondent radio bearer setup between the UE and eNB. At the eNB side, D2D flow coming from UE1 on the radio bearer RB1 are linked with the D2D flow coming from UE2 on the radio bearer RB2. Thus, the eNB maintain a layer 2 routing table in which D2D flows are associated with their correspondent radio bearers.

After the setup of the dedicated bearer between the eNB and each UE, a notification of acceptance message is sent back to the UEs to inform them about the radio resources allocated for data transmission through the eNB.

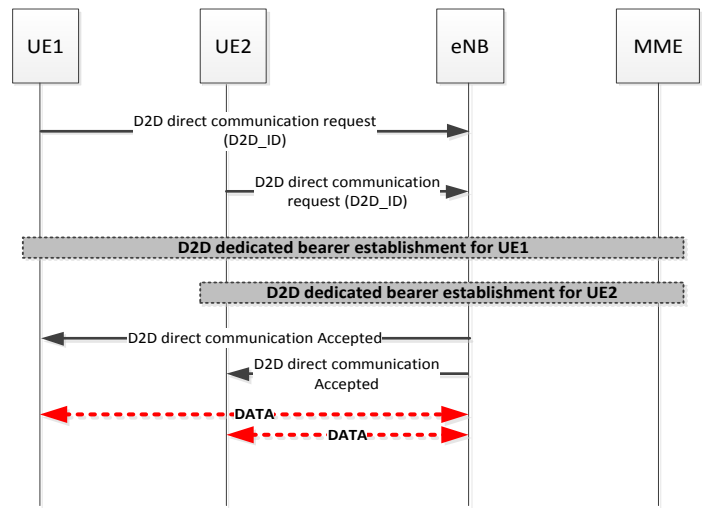

**Fig. 4.** eNB assisted D2D communication establishment

## 5 Discussion

The hybrid mechanism proposed above brings new features compared to existing D2D solutions proposed in literature. First, the direct discovery approach is combined with a network-assisted communication establishment approach. The reason of this combination is that operator networks have an important role to play in the deployment of D2D-based services as trusted parties which can satisfy users' concerns about security and privacy. It is then proposed a control plane signaling for D2D authentication and authorization during the initial attach to the network: this SSO based mechanism reduces the number of exchanged messages between the UE and the EPC network. From a long term architecture point of view, we proposed to evolve the current eNB functionalities to support D2D functions: including D2D mechanisms at the EPC level would be contradictory to the basic concept of D2D communications, which is offloading D2D traffic to local communications. One of the main objectives of our approach is to take down the D2D functionality to the lowest network entity, i.e. the closest entity to UEs. Moreover, we also propose a dedicated bearer establishment for a D2D communication using an optimized path through the eNB. As such, a service-adapted QoS could be applied through the proposed dedicated bearer establishment mechanism.

## 6 Conclusion

This paper proposes a long term operator-assisted D2D communication mechanism through a trusted procedure for a Direct D2D discovery over E-UTRA. The operator role in assisting the D2D authentication and authorization of the D2D users is highlighted. Then, a dedicated and optimized signaling for D2D communication establishment was proposed with a specific mechanism to apply a per-service QoS. In a more global view, the solution envisions a more active role of the eNB in the EPS architecture in order to meet emerging local and proximity services deployment requirements. Ongoing simulation and modelization work are carried out in order to evaluate and compare the performance of the proposed solution in terms of delays and estimate the impact of the D2D load on RAN and EPC entities.